# Geometrically frustrated magnetic behavior of $Sr_3NiRhO_6$ and $Sr_3NiPtO_6$

Niharika Mohapatra, Kartik K Iyer, Sudhindra Rayaprol[*] and E.V. Sampathkumaran[†]

*Tata Institute of Fundamental Research, Homi Bhabha Road, Colaba, Mumbai 400005, India*

The results of ac and dc magnetic susceptibility ($\chi$), isothermal magnetization and heat-capacity (C) measurements as a function of temperature (T) are reported for $Sr_3NiRhO_6$ and $Sr_3NiPtO_6$, containing magnetic chains arranged in a triangular fashion in the basal plane and crystallizing in $K_4CdCl_6$-derived rhombohedral structure. In the case of the Rh compound, the $\chi$ data reveal that there are two magnetic transitions, one in the range 10 -15 K and the other appearing as a smooth crossover near 45 K, with a large frequency dependence of ac $\chi$ in the range 10 to 40 K; in addition, the features in C(T) are smeared out at these temperatures. The magnetic properties are comparable to those of previously known few compounds with 'partially disordered antiferromagnetic structure'. On the other hand, for $Sr_3NiPtO_6$, there is no evidence for long-range magnetic ordering down to 1.8 K despite large value of paramagnetic Curie temperature. (Note: This version is an improvement over the PRB version published recently).

I. **Introduction**

The 'geometrically frustrated magnetism' is one of the important topics of research in modern condensed matter physics. While the pyrochlores and Kagome lattices have been at the center-stage of this direction of research, the compounds of the type, $(Ca,Sr)_3XYO_6$, (X,Y= transition metal ions) crystallizing in $K_4CdCl_6$-derived rhombohedral structure (space group: $R\bar{3}c$), have started attracting attention in recent years [see, for instance, Refs. 1-11]. These compounds provide a unique opportunity to probe the interplay between spin-chain phenomena as well as topologically frustrated magnetism in a single family due to triangular arrangement of antiferromagnetically coupled magnetic chains. During last few years, we have been intensely studying these compounds and found many interesting magnetic anomalies [8-11]. Among these compounds, the Co containing ones [6-11], in particular, $Ca_3Co_2O_6$, have been of considerable interest, as some of these apparently are characterized by a rare magnetic structure, called 'partially disordered antiferromagnetic (PDA) structure' [7]. In this PDA structure, two-thirds of magnetic chains at the vertices of the triangle are antiferromagnetically-coupled, whereas the third chain remains incoherent or leads to complex magnetism and spin dynamics. The readers may see some of the recent articles cited above for further exotic anomalies of these Co compounds. However, extensive studies on other compounds of this family not containing Co are still lacking and it is of interest to intensify studies in this direction; in particular, signatures of PDA structure have been rarely reported barring few claims on other isostructural compounds, $Sr_3NiIrO_6$ (Ref. 2) and $Sr_3HoCrO_6$ (Ref. 3). With this in mind, we have subjected $Sr_3NiRhO_6$ (Ref. 4) and $Sr_3NiPtO_6$ (Ref. 5) to detailed ac and dc magnetic susceptibility ($\chi$), isothermal magnetization (M) and heat capacity (C) studies. While the former appears to belong to 'PDA' family [12] considering similarities in the magnetic behavior of $Ca_3CoRhO_6$, the latter does not show any signature of long range magnetic ordering down to 1.8 K.

## II. Experimental details

The polycrystalline samples of $Sr_3NiRhO_6$ and $Sr_3NiPtO_6$ were prepared via a solid state route starting from powders of $SrCO_3$ (99.999%), NiO (99.999%), Rh (>99.9%) and $PtO_2$ (99.999%). For $Sr_3NiRhO_6$, the intimately mixed powders of the constituent components in proper proportion were calcined at 800 C for 1 day and the material in the form of a pellet was subsequently sintered at 900 C for 1 day, 1000 C for 1 day and 1150 C for 30 hours (3 times) with intermediate grindings. In the case of Pt sample, after initial calcination, the sintering of the pellet was done at 1000 C for 9 days with three intermediate grindings. The samples were characterized by x-ray diffraction and found to be single phase (Fig. 1). The lattice constants (± 0.004 Å) are found to be in good agreement with those reported in the literature [4,5]: $Sr_3NiRhO_6$, $a$ = 9.588 Å, $c$ = 11.061 Å; $Sr_3NiPtO_6$, $a$ = 9.583 Å, $c$ = 11.155 Å. In addition, the composition and the homogeneity were confirmed by scanning electron microscope. Dc $\chi$ measurements (T = 1.8 – 300 K) were performed by commercial magnetometers (superconducting quantum interference device, Quantum Design, USA, and vibrating sample magnetometer, Oxford Instruments, UK) in the presence of few magnetic fields (H= 100 Oe, 1, 5, 10 and 100 kOe). Isothermal M behavior at selected temperatures were tracked with the vibrating sample magnetometer up to 120 kOe. In addition, ac $\chi$ measurements (1.8 – 75 K) were performed with four frequencies ($\nu$= 1.3, 13.3, 133 and 1300 Hz; $H_{ac}$= 1 Oe). Heat capacity data were collected by relaxation method with the help of a commercial physical property measurements system (Quantum Design, USA).

## III. Results and discussions
### A. $Sr_3NiRhO_6$
**Magnetic susceptibility:**

The results of dc $\chi$ measurements for $Sr_3NiRhO_6$ for zero-field-cooled (ZFC) and field-cooled (FC) conditions of the specimen are shown in figure 2. For ZFC-condition, there is a gradual increase of $\chi$ with decreasing temperature down to about 15 K, below which there is a sudden fall attributable to magnetic ordering; however FC-curves exhibit an increase without any fall; the magnetic ordering in the FC-curve is manifested in the form of a kink near 12 K. The $\chi$ for both the curves undergoes a weak increase below 10 K. These findings are in agreement with Ref. 4. However, we note important features which were not reported in the earlier literature. That is, (i) we observe a broad shoulder in $\chi(T)$ around 50 - 70 K, and below 45 K, there is a change increase in slope, as shown in figure 3 for H= 5 kOe. This upturn gradually gets diminished with increasing field; if a high field is applied, say, 120 kOe, this upturn is absent. Thus, there is a suppression of $\chi$ with increasing fields in the intermediate temperature range, say 10 to 45 K. (ii) The temperature at which ZFC-FC curves bifurcate decrease with increasing fields; for instance, it occurs near 40 and 20 K for H= 100 Oe and 10 kOe (see figure 2). These features are qualitatively in agreement with those observed [10] for $Ca_3CoRhO_6$, which has been established to undergo PDA ordering in the range 30 - 90 K; and (iii) the ZFC-FC curves nearly overlap for all fields (though marginally lower for H=100 kOe) at low temperatures, as in $Ca_3CoRhO_6$ when the third chain undergoes spin-glass-like freezing (below 30 K). Therefore, we propose that, as the temperature is lowered, in the Ni compound, there is a magnetic transition near (T1= ) 45 K partially appearing as a 'smooth crossover' from paramagnetism (rather than as a sharp feature possibly due to some disorder), and the feature around (T2= ) 10-15 K (proposed in Ref. 4 to be the only magnetic transition) is actually a second transition. The broad shoulder regime (50 – 70 K) apparently arises from short range magnetic fluctuations by analogy with $Ca_3CoRhO_6$ (Ref. 10).

For the sake of comparison with Ref. 4, before we proceed on further discussions on T1 and T2, we look at the data in the paramagnetic state. Above 150 K, $\chi(T)$ follows Curie-Weiss law (see figure 3). The value of the effective moment obtained from the high temperature linear region turns out to be 3.3 $\mu_B$ per formula unit. This agrees well with the theoretical spin-only moment for $Rh^{4+}$ in the octahedral coordination ($4d^5$, low-spin) and $Ni^{2+}$ ($3d^8$) in the trigonal prismatic site, in agreement with Ref. 4. Below 150 K, however, a gradual deviation from the high temperature linearity is noticed due to gradual dominance of magnetic short-range correlations. The value of the paramagnetic Curie-temperature ($\theta_p$) turns out to be about -24 K comparable to, say, T1 or T2 (unlike in pyrochlores in which case $\theta_p$ values are comparatively larger), with the negative sign indicating strong antiferromagnetic correlations,

**Isothermal magnetization:**

In order to gather support for the above interpretation of magnetic properties, we have compared isothermal M behavior in the relevant temperature ranges (see Fig. 4). At 100 K, M varies linearly with H, typical of paramagnets, and at 75 K, a curvature develops beyond about 80 kOe as though there is a tendency for saturation at much higher fields. This feature endorses our view that short range magnetic correlations are present well above T1. As the temperature is lowered to 40 K, that is, in the T-range between T1 and T2, there is a deviation from the linearity, resulting in a clear-cut curvature beyond 60 kOe as though there is a spin-reorientation. The non-linearity (below 40 kOe) and this curvature are quite prominent at 25 K. These curves are non-hysteretic. This implies that the zero-field state is essentially antiferromagnetic. As one enters the region below T2, say at 5 K, the feature attributable to spin-reorientation is absent and the variation of M with H is rather sluggish at higher fields (when compared to low-field linear extrapolation), however with an observable hysteresis. Absence of a tendency for saturation and hysteretic behavior at this temperature viewed together imply coexistence of antiferromagnetic and ferromagnetic coupling, thereby pointing towards spin-glass freezing. These M(H) curves thus reveal in any case that there are two distinct temperature ranges with different zero-field magnetic structures, supporting our conclusions earlier in this article.

We now offer arguments for PDA structure between T1 and T2 on the basis of magnetization data. As already remarked, in the PDA structure, two out of three chains are anti-parallel to each other with the third remaining incoherent. As the incoherent chain gets oriented with the application of magnetic field, there is a net magnetization of 1/3 of the saturation magnetization at a certain H value. In polycrystalline samples [3], this usually appears as a tendency for a step or a sudden increase in the slope in the M(H) plot as a characteristic feature of PDA structure. The M(H) plots at 25 and 40 K, closely resembling those in $Sr_3NiIrO_6$ (Ref. 2) as well, reveal that such a feature occurs in the present case in the field range 60-70 kOe near 0.3 $\mu_B$ per formula unit. In order to infer whether it is consistent with PDA, it is important to know the value of the saturation magnetization. But it is obvious from figure 4 that there is no evidence for saturation even at high fields. If one assumes that intra-chain coupling among Ni (S= 1) and Rh (S= ½) spins is ferromagnetic, one would expect a saturation moment of 3 $\mu_B$ per formula unit, which is then not in agreement with the step-value. If we have to assume that the intra-chain Ni-Rh interaction is antiferromagnetic, a saturation magnetization value of 1 $\mu_B$ per formula unit is expected, which is then consistent with a possible step around 0.3 $\mu_B$. Considering that the sign of $\theta_p$ is negative, this proposal may be true. For comparison, in the case of $Ca_3CoRhO_6$, in which the intra-chain interaction is ferromagnetic [7], the sign of $\theta_p$ is positive (175 K, Ref. 10). In short, the results could be interpreted consistently in terms of PDA structure between T1 and T2, if we assume that intra-chain interaction is antiferromagnetic. In this sense (that is, intrachain antiferromagnetic coupling), this compound is different from other members of this family and comparable to the behavior of Cs systems (see, Mekata, Ref. 12). Looking at the nature of M(H) curve at 5 K, PDA structure is apparently destroyed below T2.

**Ac susceptibility:**

It may be recalled that the compounds exhibiting PDA structure show interesting spin dynamics as revealed by an unusually large ν-dependence of ac susceptibility [10, 11]. In order to explore this in the present case, we have performed ac χ measurements. It is obvious from figure 5 that there is a peak in the real part ($\chi'$) of ac χ in the range 18-21 K for ν= 1.3 Hz, which is few degrees higher than that in the dc χ in figure 2, whereas the peak in the imaginary part ($\chi''$) occurs at 15 K. This trend in the peak positions of $\chi'$ and $\chi''$ with respect to dc χ is similar to that observed for $Ca_3CoRhO_6$ and $Sr_3NiIrO_6$ (Refs. 2, 10, 11). With increasing ν, the curves undergo a shift to higher temperatures as in spin glasses, but the magnitude of the shift is very large compared to canonical spin glasses. For instance, the curves shift by about 3.5 K as ν is increased to 133 Hz from 1.3 Hz resulting in a value of about 0.1 for the factor $\Delta T_f/T_f\Delta(\log \nu)$ (Ref. 13), as in other compounds [2,10], if one assumes that the peak temperature corresponds to $T_f$. As shown in the inset of figure 5, interestingly, Vogel-Fulcher formula, $\nu = \nu_0 \exp[-E_a/k_B(T_f-T_a)]$ (where $E_a$ is the activation energy and $T_a$ is considered to be a measure of intercluster interaction strength) usually obeyed in dilute spin-glass systems, is found to be valid in this case as well. Assuming that $\nu_0$ to be the same as in spin-glasses [13], the value of $E_a$ is determined to be 155 K. We have also performed these studies in the presence of 10 and 70 kOe; a trend in ν-dependence can be inferred even at such fields (see, for instance, H= 10 kOe data), though the data is noisy.

**Isothermal remnant magnetization:**

We have also obtained isothermal remanent magnetization ($M_{IRM}$) behavior at 1.8 and 30 K. For this purpose, we cooled the sample in zero field to desired temperature, switched on a field of 5 kOe for 5 mins, following which the field was switched off and the data was collected as a function of time (*t*). We notice that (see figure 4. inset) $M_{IRM}$ undergoes slow relaxation with time varying logarithmically with *t*, but not stretched-exponentially noted for $Sr_3NiIrO_6$ (Ref. 2). The coefficient of logarithmic term is comparable to that in $Ca_3CoRhO_6$. This behavior below T1 can be described in terms of 'spin-glass'. At 30 K, $M_{IRM}$ fell to negligibly small values within few minutes after the field is switched off, thereby suggesting that there are two distinctly different magnetic regions as the temperature is varied and that the third chain is truly paramagnetic-like and not spin-glass-like between T1 and T2.

**Heat-capacity:**

We show heat capacity behavior as a function of temperature in figure 6. There is no evidence for any λ-anomaly attributable to magnetic ordering in the temperature range of investigation (1.8 to 75 K)**.** The plot of C/T is also featureless. We find that C/T tends to a constant value (~ 80 mJ/mol $K^2$) below 5 K and the linear term in the range 5 to 10 K is also quite large despite the fact that this is an insulator ((about 65 mJ/mol $K^2$) (see figure 6, inset). Such a large linear term is consistent with spin-glass freezing below 10 K [14]. The Debye temperature is estimated to be 290 K from the data in the range 5 to 10 K. The fact that the feature near T1 is also smeared out implies that there is a gradual loss of entropy due to short range correlations within the chains at higher temperatures [15].

### B. $Sr_3NiPtO_6$

The results of $\chi$, isothermal M and C measurements are shown in figure 7 for this compound. We do not find any difference between ZFC and FC $\chi(T)$ curves and no feature in ac $\chi$ could be observed down to 1.8 K. These results rule out any kind of magnetic ordering including spin-glass freezing in the temperature range of investigation. The M(H) plots are practically linear up to 120 kOe (but for a weak curvature at low-fields as though there is a small ferromagnetic component) and non-hysteretic, thereby confirming that the material is essentially paramagnetic. At this juncture, it is to be mentioned that a previous article (Ref. 5) reports a sudden change in M value near 10 kOe in an oriented specimen as though there is a field-induced ferromagnetism. We do not see any signature of such a sudden induced ferromagnetism in polycrystals, possibly because the intensity of this feature (Ref. 5) is rather weak. The features in $\chi(T)$ are essentially the same as that reported by Vajenine et al [4] for the polycrystalline sample in the sense there is a monotonic increase of $\chi$ with decreasing temperature, with a tendency to flatten below about 30 K as often observed in low dimensional materials. There is an upturn below 10 K due to ferromagnetic impurities and if the values are determined from the slope of M-H plot above 60 kOe, we could observe the low temperature flattening.. The $\chi(T)$ exhibits Curie-Weiss behavior above 150 K, below which there is a gradual deviation from this behavior presumably due to short range magnetic correlations. The value of $\mu_{eff}$ and $\theta_p$ (above 200 K) are found to be nearly 3.4 $\mu_B$ and -25 K. The negative sign of $\theta_p$ implies dominant anti-ferromagnetic correlations. The magnitude of $\mu_{eff}$ is larger than the spin-only value for high spin $Ni^{2+}$ ion (2.83 $\mu_B$). Since Pt is believed not to possess a magnetic moment [4] (as it is in the low spin 4+ state), the origin of excess moment is not clear to us at present. With respect to the heat-capacity behavior (see figure 7b), there is neither a $\lambda$-anomaly nor any shoulder in C(T) in the range 1.8 – 75 K. We have not attempted to deduce magnetic contribution to C, as there is no reliable reference compound for lattice contribution. In any case, the results reveal that there is no evidence for long range magnetic ordering down to 1.8 K. It is not clear whether 'spin-liquid' phenomenon is partly responsible for this behavior, interfering with possible single-ionic effects proposed earlier [5]. Future experiments may throw more light on this aspect. We would like to emphasize that the data taken in the presence of a high magnetic field, say 50 kOe, does not show up any new feature in C(T) and the in-field curve overlaps with that of zero-field. This could be against any field-induced ferromagnetism in this compound. A notable observation is that C/T undergoes marginal upturn below 6 K, with a tendency to flatten below 2.5 K similar to $\chi(T)$ behavior, to a non-zero value (25 mJ/mol $K^2$), similar to Fermi liquids, implications of which is not clear at present. Finally, the Debye temperature obtained from the data in the region 10 to 20 K turns out to be 300 K.

### IV. Conclusions

We have subjected $Sr_3NiRhO_6$ and $Sr_3NiPtO_6$ to detailed magnetic investigations. The compound $Sr_3NiRhO_6$ is characterized by a magnetic crossover near 45 K and another more well-defined magnetic transition near 10 K. We have compared and contrasted its magnetic properties with those of $Ca_3CoRhO_6$ and $Sr_3NiIrO_6$ and we infer that the magnetic behavior is typical of PDA structure between 10 and 45 K; below 10 K, spin-glass freezing is proposed;

complex spin dynamics may be operative as revealed by ac susceptibility. It therefore appears that the presence of Co at the trigonal prismatic site is not crucial for PDA-like anomalies. On the other hand, the compound $Sr_3NiPtO_6$ does not undergo long range magnetic ordering down to 1.8 K. Interestingly, in this compound, C/T and $\chi$ tend to a large and constant value at low temperatures. Under the assumption that Pt does not carry any moment, similar magnetic properties of Rh and Ir compounds (among these Sr compounds) indicate that the direct d-d overlap, rather than super-exchange paths involving oxygen, may be a crucial factor to result in PDA structure. We hope that this serves as a hint for further theoretical understanding of geometrically frustrated magnetism in this class of spin-chain compounds. Finally, we have not attempted to determine intra-chain and inter-chain interaction strengths, as it is not easy to make a realistic estimate of these due to various difficulties as discussed in Ref. 16. It will be fruitful to perform other microscopic studies like neutron diffraction, $^{195}$Pt NMR, and μSR studies in these compounds.

We thank Kedar Damle and Vijay Shenoy for fruitful discussions.

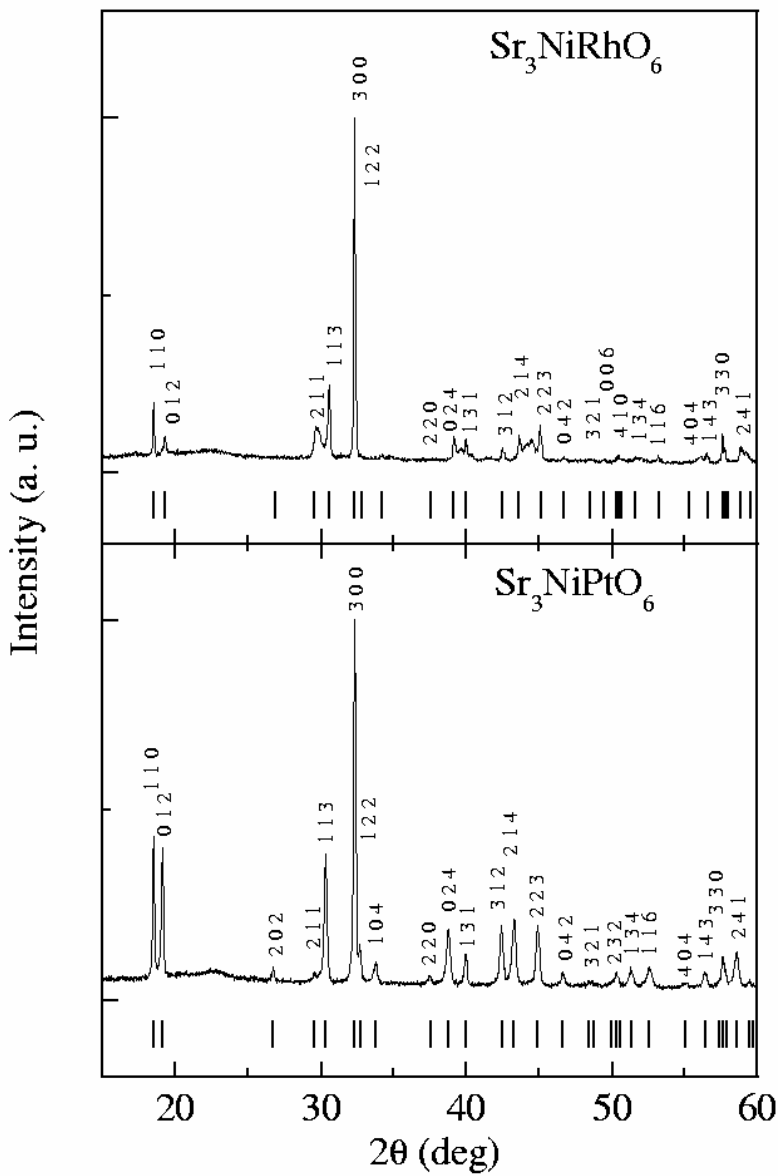

Figure 1

(color online) X-ray diffraction patterns (Cu $K_\alpha$) of $Sr_3NiRhO_6$ and $Sr_3NiPtO_6$. The vertical bars represent expected peak positions.

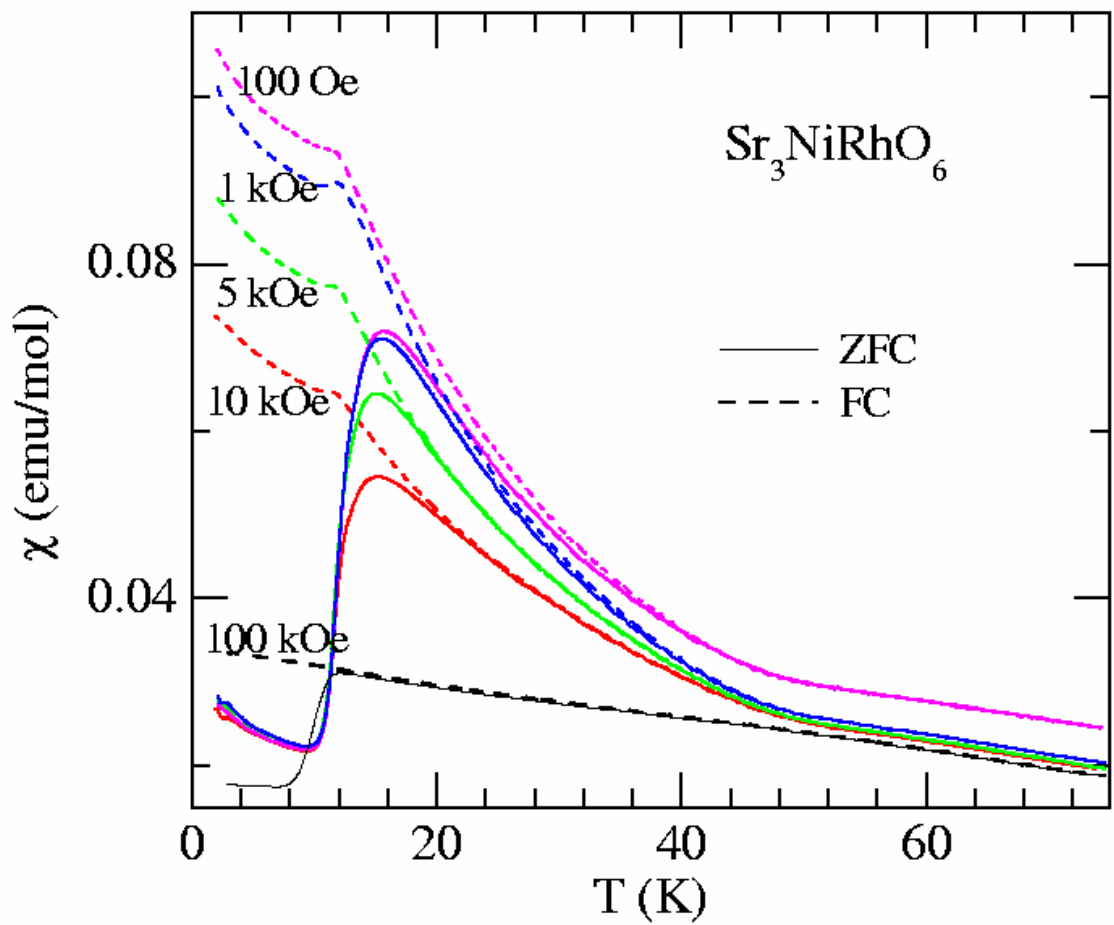

Figure 2

(color online) Dc magnetic susceptibility as a function of temperature (below 75 K) for the zero-field-cooled (ZFC, continuous line) and field-cooled (FC, discontinuous line) conditions of the specimen of $Sr_3NiRhO_6$ measured in the presence of various magnetic fields.

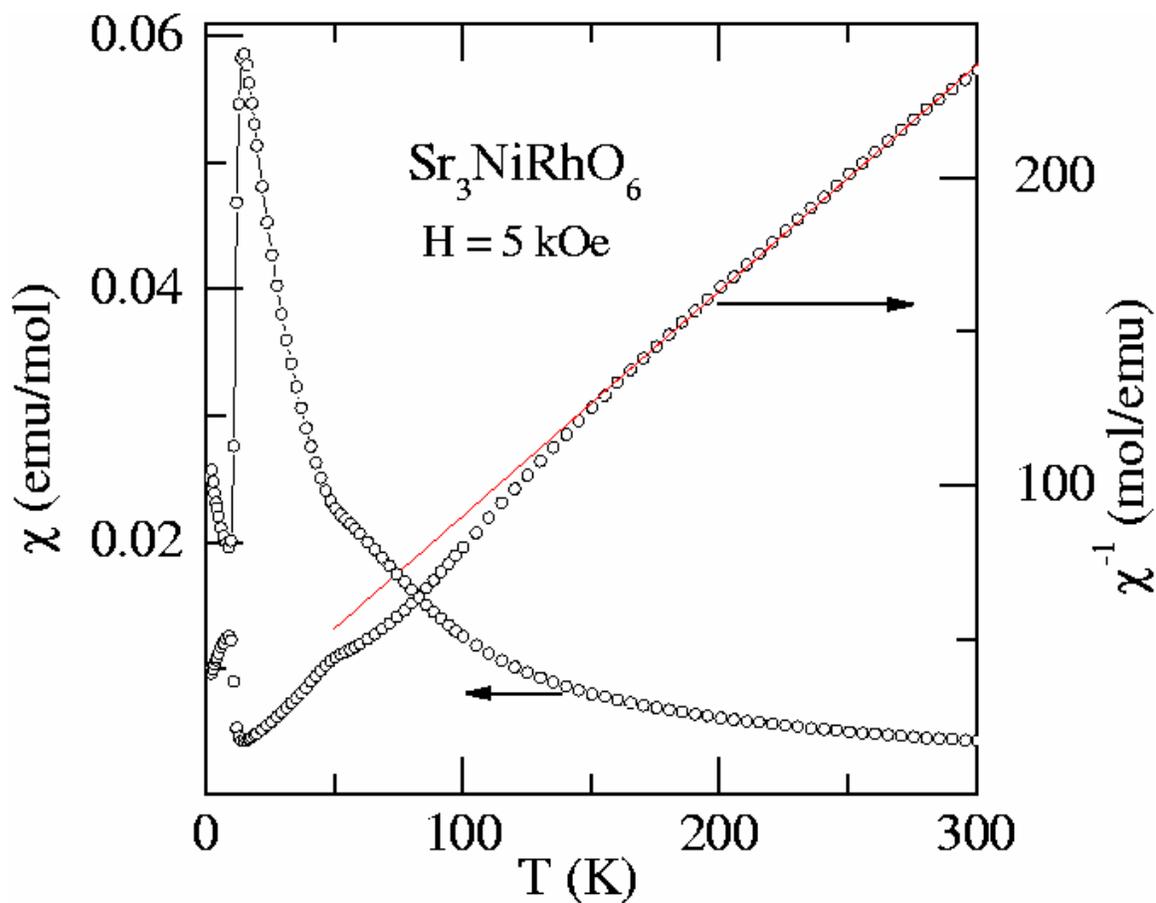

Figure 3

(color online) Dc magnetic susceptibility as a function of temperature (1.8 – 300 K) for the zero-field-cooled (ZFC) condition of the specimen of $Sr_3NiRhO_6$ measured in the presence of 5 kOe. The continuous line in the plot of $\chi^{-1}$ represents least-square fitting of the data in the linear region above 150 K, whereas, in the plot of $\chi(T)$, the line is a guide to the eyes.

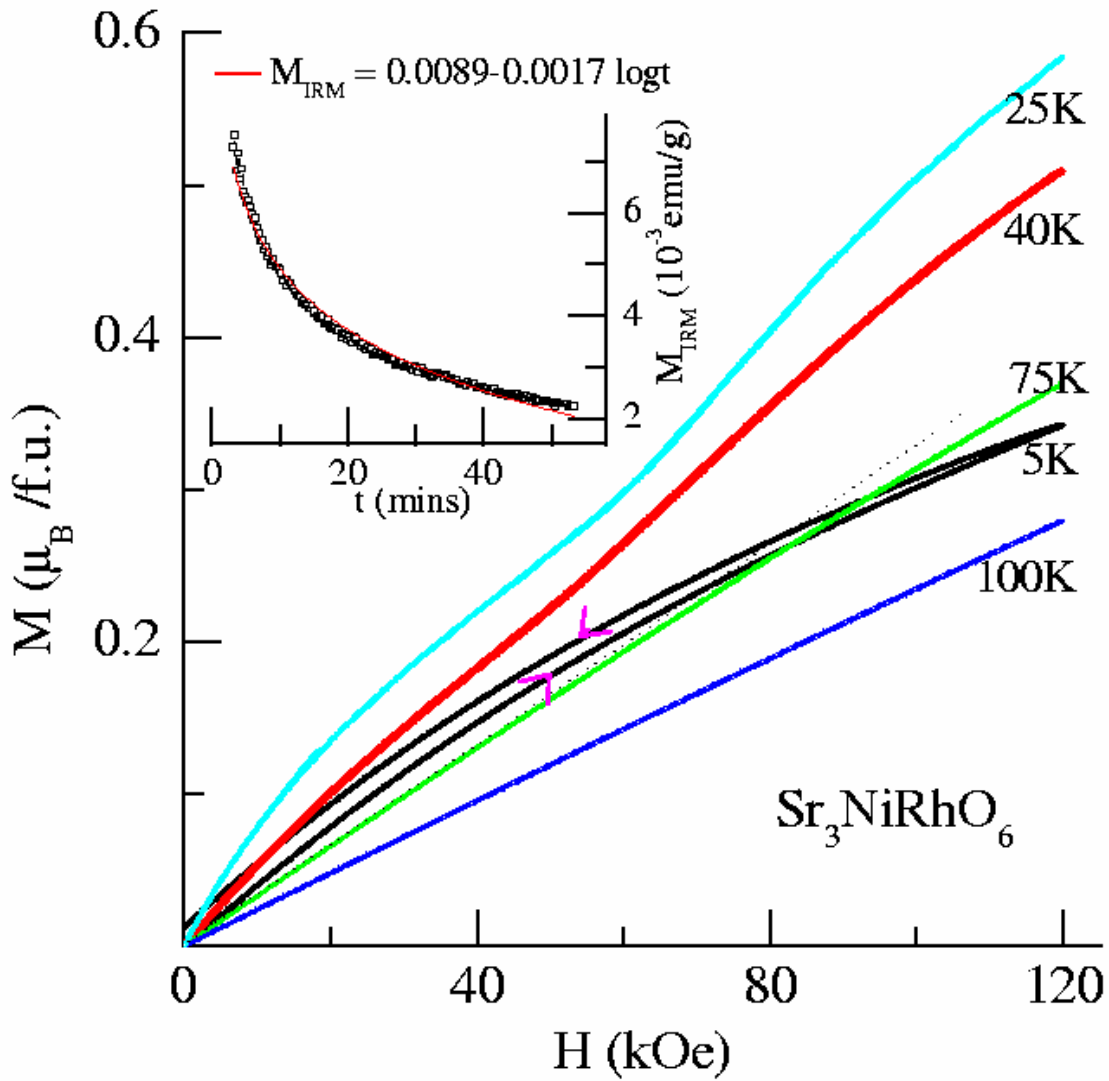

Figure 4

(color online) Isothermal dc magnetization behavior of $Sr_3NiRhO_6$ at selected temperatures. The curves for up and down field-cycling overlap for all temperatures (except for 5 K, in which case there is a hysteresis). Inset shows isothermal remnant magnetization behavior as a function of time at 5 K. The dashed line for 75K-data is obtained by linear extrapolation of the low-field data.

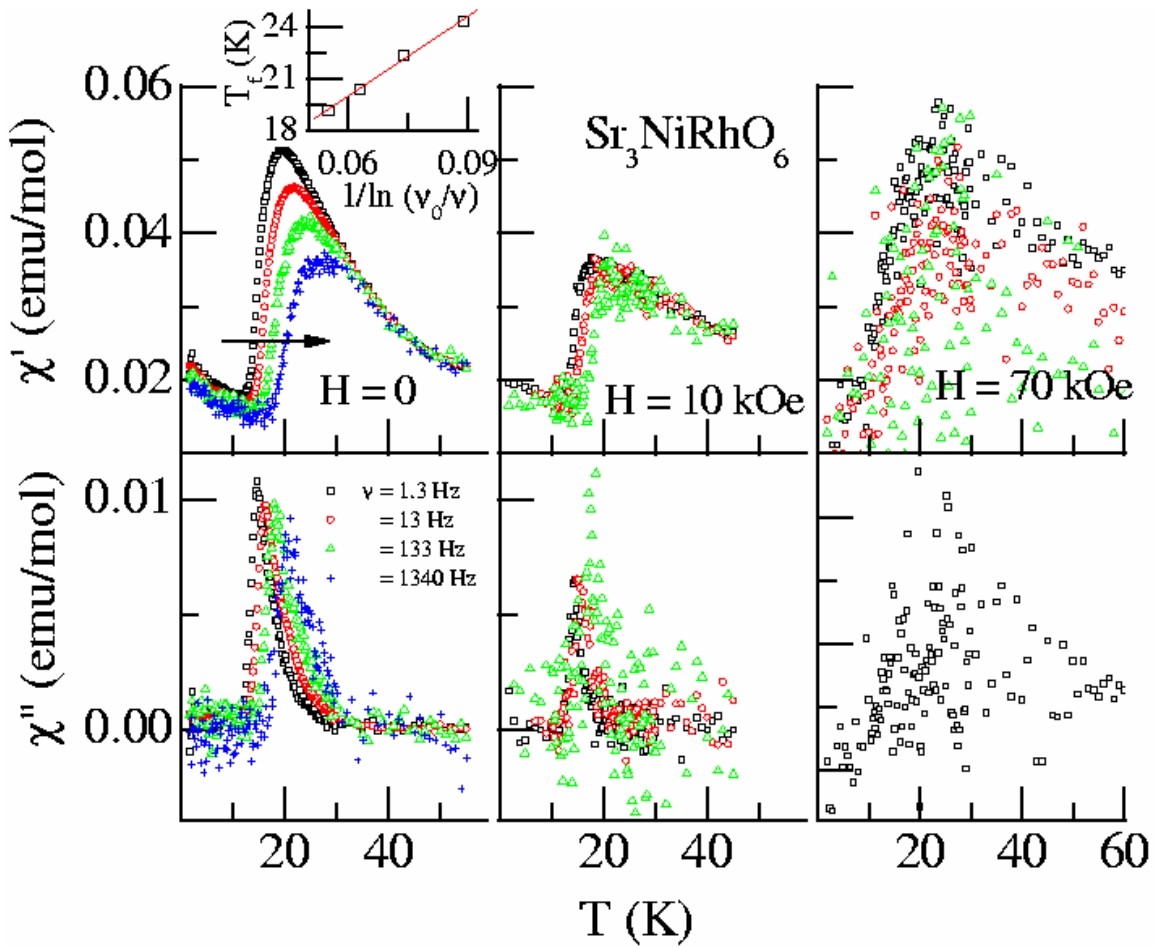

Figure 5

(color online) Real ($\chi'$) and imaginary ($\chi''$) parts of ac magnetic susceptibility at various frequencies as a function of temperature for $Sr_3NiRhO_6$. The arrow indicates the direction in which the curves move with increasing frequency. The inset shows a plot demonstrating validity of Vogel-Fulcher relationship for the H= 0 data. The data at higher fields are very noisy and the ones at higher frequencies have been omitted for the sake of clarity.

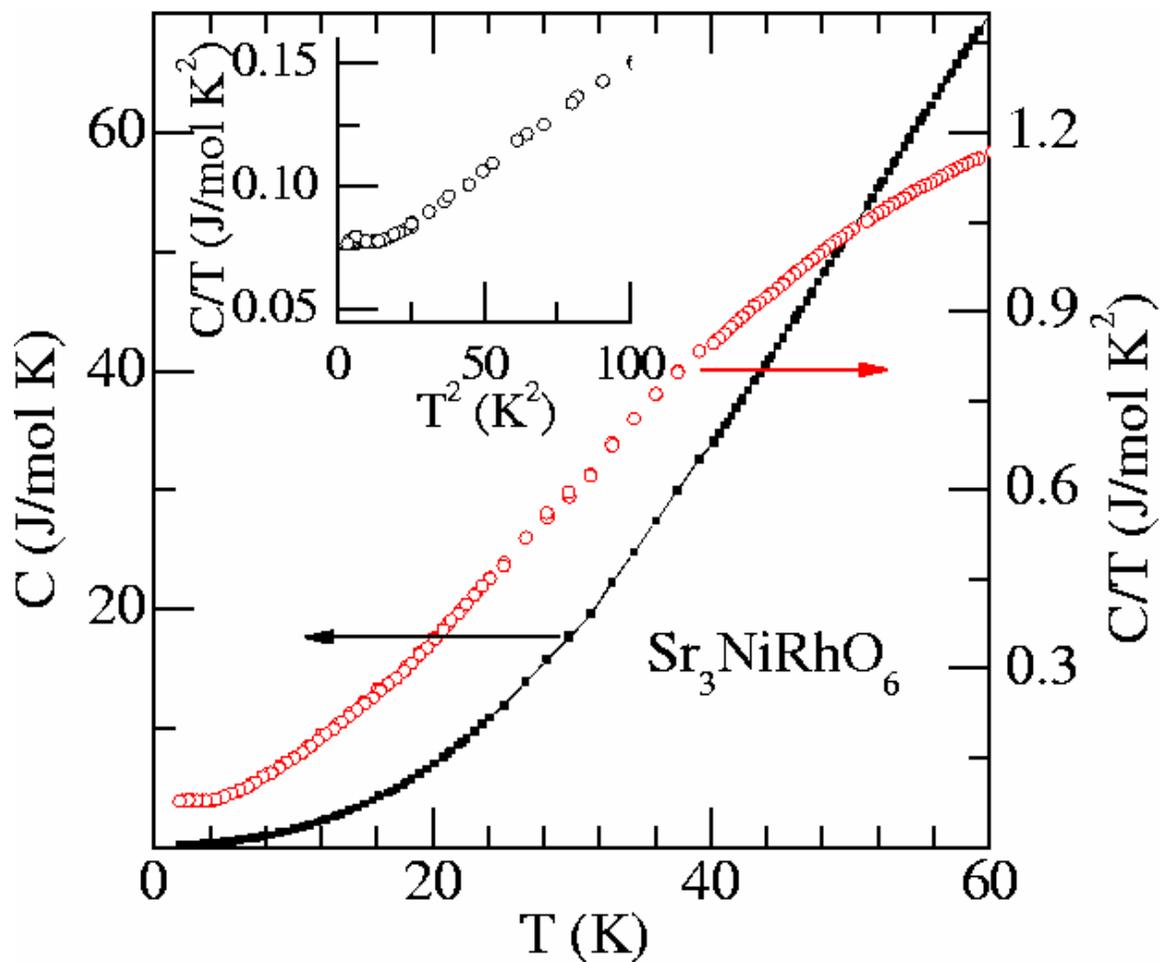

Figure 6
(color online) Heat capacity (H= 0) as a function of temperature for $Sr_3NiRhO_6$ plotted in various ways. The continuous line through the data points is a guide to the eyes.

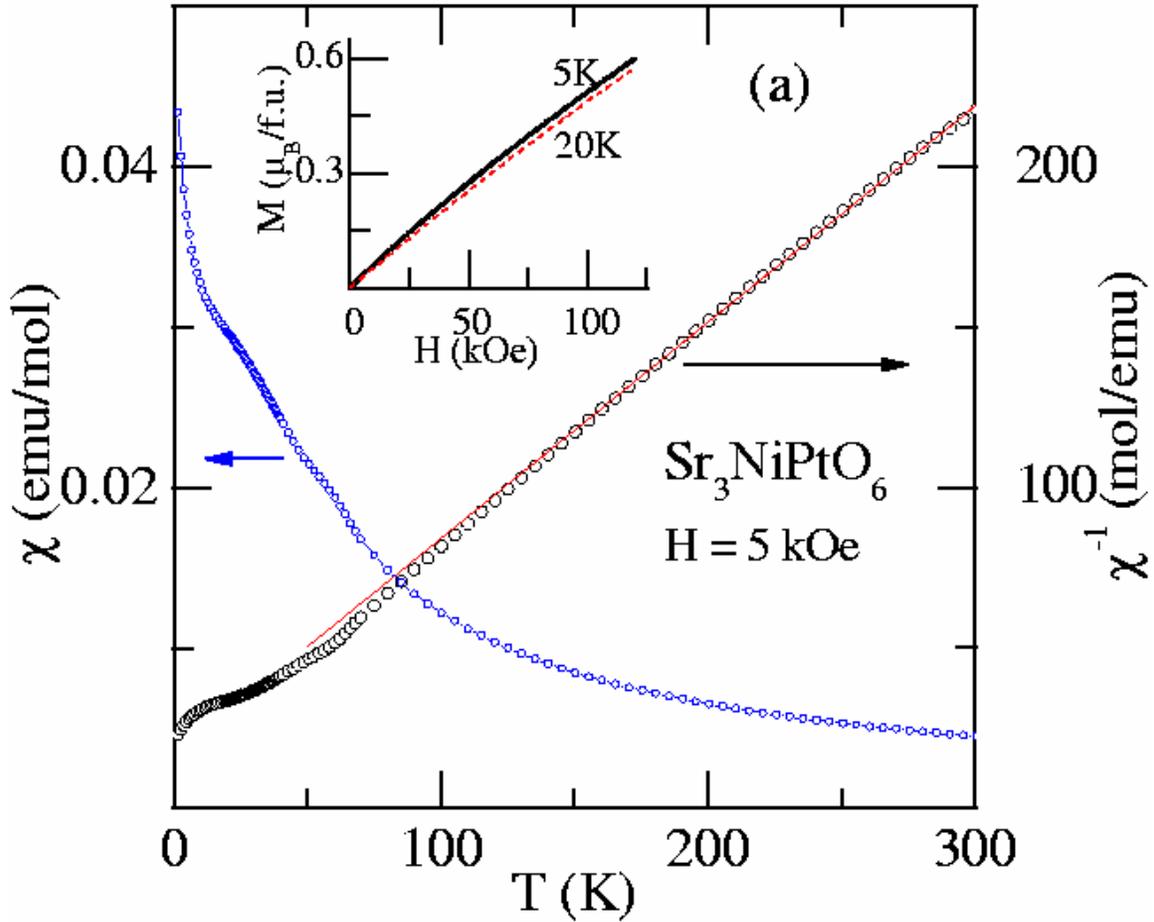

Figure 7a
(color online) Dc magnetic susceptibility as a function of temperature for $Sr_3NiPtO_6$ obtained in a field of 5 kOe, plotted in different ways. The continuous line in the plot of inverse $\chi$ versus T represents high temperature Curie-Weiss region, whereas, in the $\chi(T)$ plot, the line is a guide to the eyes. The inset shows isothermal magnetization data at 5 and 20 K.

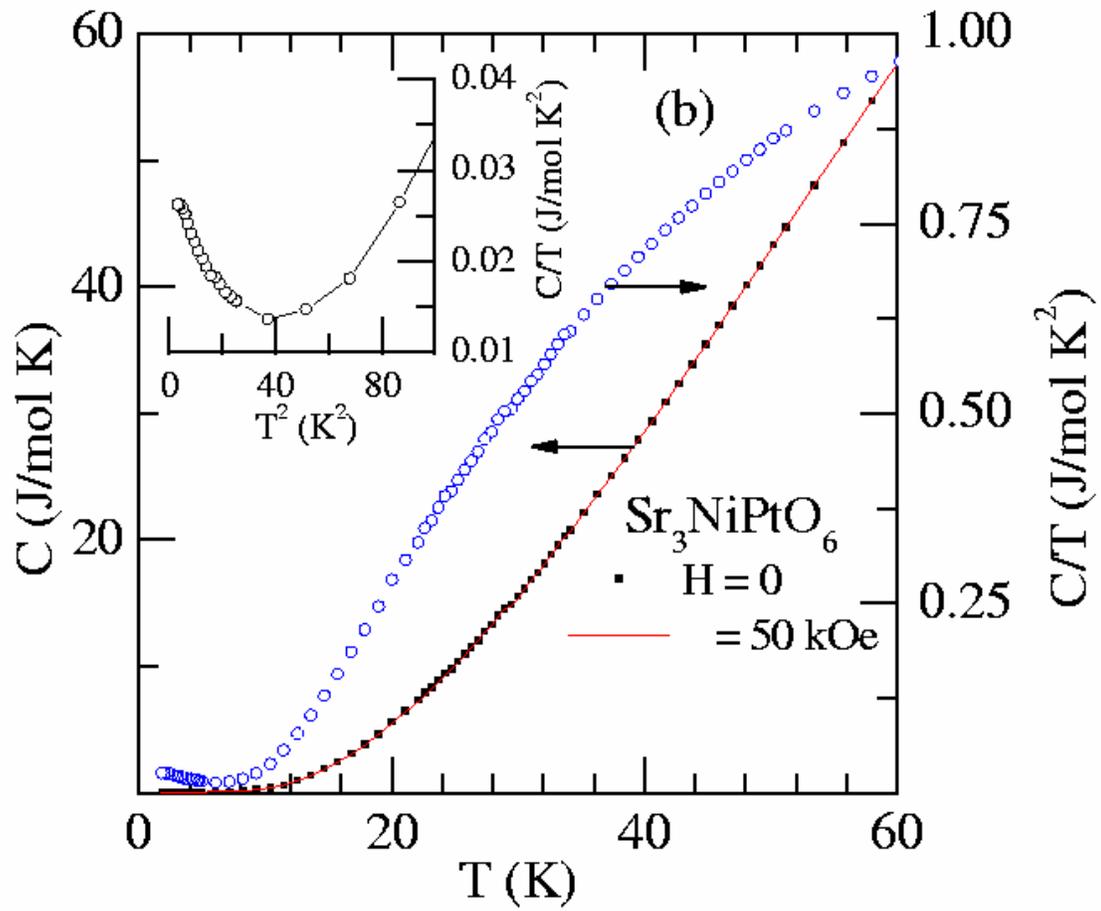

Figure 7b

(color online) Heat capacity (C) and C/T as a function of temperature (T) for $Sr_3NiPtO_6$. In the inset, the data at low temperatures is plotted as a function of $T^2$.